\documentclass[conference]{IEEEtran}
\IEEEoverridecommandlockouts
\usepackage{cite}
\usepackage{amsmath,amssymb,amsfonts}
\usepackage{algorithmic}
\usepackage{graphicx}
\usepackage{textcomp}
\usepackage{xcolor}
\def\BibTeX{{\rm B\kern-.05em{\sc i\kern-.025em b}\kern-.08em
    T\kern-.1667em\lower.7ex\hbox{E}\kern-.125emX}}

\ifCLASSOPTIONcompsoc
\usepackage[caption=false,font=normalsize,labelfon
t=sf,textfont=sf]{subfig}
\else
\usepackage[caption=false,font=footnotesize]{subfig}
\fi

\begin{document}

\title{Knowledge Federation: A Unified and Hierarchical
Privacy-Preserving AI Framework}



\author{
	\IEEEauthorblockN{Hongyu Li*, Dan Meng, Hong Wang and Xiaolin Li}
	\IEEEauthorblockA{\textit{AI Institute, Tongdun Technology}\\
		Shanghai, China\\
		\{hongyu.li, dan.meng, hong.wang, xiaolin.li\}@tongdun.net}
	}

\maketitle

\begin{abstract}
With strict protections and regulations of data privacy and security, conventional machine learning based on centralized datasets is confronted with significant challenges, making artificial intelligence (AI) impractical in many mission-critical and data-sensitive scenarios, such as finance, government, and health. In the meantime, tremendous datasets are scattered in isolated silos in various industries, organizations, different units of an organization, or different branches of an international organization. These valuable data resources are well underused. To advance AI theories and applications, we propose a comprehensive framework (called Knowledge Federation - KF) to address these challenges by enabling AI while preserving data privacy and ownership. Beyond the concepts of federated learning and secure multi-party computation, KF consists of four levels of federation: (1) information level, low-level statistics and computation of data, meeting the requirements of simple queries, searching and simplistic operators; (2) model level, supporting training, learning, and inference; (3) cognition level, enabling abstract feature representation at various levels of abstractions and contexts; (4) knowledge level, fusing knowledge discovery, representation, and reasoning. We further clarify the relationship and differentiation between knowledge federation and other related research areas. We have developed a reference implementation of KF, called iBond Platform, to offer a production-quality KF platform to enable industrial applications in finance, insurance, marketing, and government. The iBond platform will also help establish the KF community and a comprehensive ecosystem and usher in a novel paradigm shift towards secure, privacy-preserving and responsible AI.  As far as we know, knowledge federation is the first hierarchical and unified framework for secure multi-party computing (statistics, queries, searching, and low-level operations) and learning (training, representation, discovery, inference, and reasoning).
\end{abstract}


\section{Introduction}

Since AlphaGo and AlphaZero defeated top human professional Go players \cite{silver2017mastering}, artificial intelligence has made leap progresses, rapidly applied in numerous scenarios and domains, such as object recognization, recommender systems, finance, precision medicine, autonomous driving, and smart city. AI has become an essential part of our daily life, work, entertainment, and digital infrastructure.

In the past decade, the rapid advancement of AI has been mainly driven by large data volume, powerful graphics processing unit (GPU), and breakthroughs in deep learning, reinforcement learning, and other machine learning techniques. Deep learning methods have proved effective in real-world applications but require powerful computing resources and a huge amount of data during training in order to prevent overfitting. Typically, to train a deep model, various transactional datasets that may involve privacy are collected from users or institutions and stored in a central server. Such approach with central data repositories is vulnerable because the central server may encounter data unsafety or privacy leakage due to advanced cyber attacks or personnel's mistakes.

Data security and privacy has attracted global attention in recent years. In 2016, China passed its first Cybersecurity Law, aiming to strengthen cyberspace governance through a number of initiatives, including personal information protection, special protection of critical information infrastructure, and local storage of data. The General Data Protection Regulation (GDPR)~\cite{voigt2017eu} took effect in 2018. GDPR has been designed to provide individuals with greater control over how their personal data is collected, stored, transferred, and used, while also simplifying the regulatory environment across the European Union (EU). In 2020, Personal Information Security Specification and Personal Financial Information Protection Technical Specification are officially announced in China. A series of lawsuits against Internet giants due to privacy issues result in tremendous capital loss for commercial companies.

\begin{figure*} [htb]
	\centering
	\includegraphics[width=\linewidth]{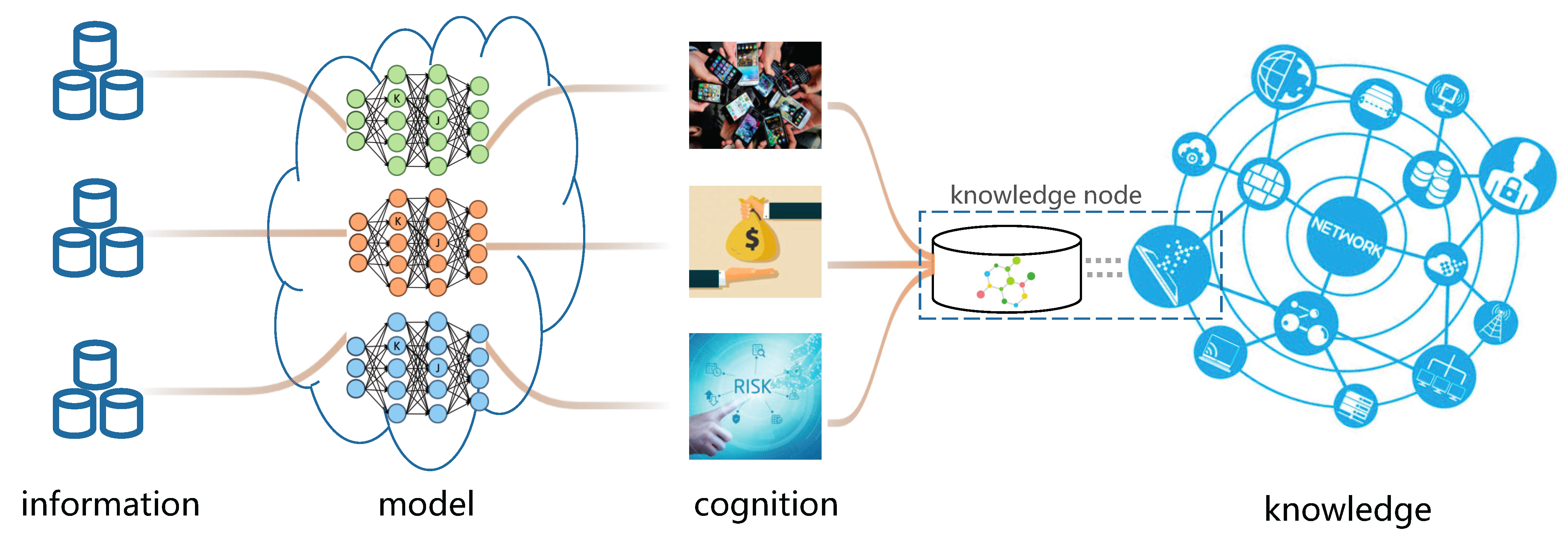}
	\caption{Hierarchy of Knowledge Federation. KF features four levels of federation: information level, model level, cognition level and knowledge level.}
	\label{fig:four-level}
\end{figure*}
It is a new challenge to discover AI knowledge from big data while not compromising data security and privacy. To address such challenges, researchers proposed a federated learning framework~\cite{mcmahan2016communication} for training a privacy-preserving model. The initial idea is to enable multiple devices to collaboratively learn a shared prediction model while all the training data is kept locally on device. A federated averaging algorithm is proposed to greatly reduce training rounds and communication overheads for converging~\cite{konevcny2016federated0} . The communication costs in one round can be further reduced by compressing gradient updates using random rotation and quantization. \cite{bonawitz2017practical} developed a secure aggregation protocol by encrypting participant's local gradients before aggregation.

Federated learning methods, however, focus more on the safe model training based on encrypted gradient updates. In fact, there exist other strategies for secure federation that are not restricted to learning from multi-party data. One solution is to train models or utilize prior knowledge in a ciphertext space on a central sever where primitive data is encrypted before pooling together and is not decrypted during training or computing. Another solution is to first extract crude knowledge from each participant, which is encoded with deep neural networks or traditional machine-learning models, and then refine knowledge through ensemble or aggregation on a server. When more and more knowledge is extracted and stored as knowledge nodes, connecting such nodes from different participants together will naturally form a knowledge network that contributes to further secure knowledge reasoning.

To unify and fuse various secure computing and learning techniques, we propose a novel framework, called knowledge federation (KF), with a coherent hierarchy of four levels of federation: information level, model level, cognition level, and knowledge level. Leveraging best practices in conventional cryptography, recent progresses in secure multi-party computation and federated learning, and rapid advancement of distributed machine learning, KF features a unified framework with salient hierarchical abstractions of four-level federation, paving the way towards the next generation of secure AI. We also briefly present a reference implementation of KF framework: iBond platform.

\section{Conceptual Architecture of Knowledge Federation}

\subsection{Concept of Knowledge Federation}
Knowledge federation, as the name indicates, is the combination of two terms, knowledge and federation. Knowledge is generally considered as valuable information related to decision making or principles acquired by humankind. In knowledge federation, knowledge refers to models learned from available data, or rules sourced from practical experience. 
Federation often means cooperation protocols among several self-governing entities under the supervision of a central administrative unit. In this case, it amonuts to secure protocols on data exchange among multiple independent data holder through a trustworthy third party.

In this work, \textit{knowledge federation} is formally defined as collaboratively creating or utilizing significant knowledge over isolated mutli-party data through secure protocols that prevent privacy leakage during data exchange. Specifically, given a set of datasets $\chi=\{\chi_{i=1\cdots n}$\}, each $\chi_i$ distributed on an independent party $P_i$, we expect to learn some useful knowledge $\kappa$ from multi-party data $\{\chi_i\}$ with known labels $y$,
\begin{equation}\label{kf_learn}
\kappa:\chi\rightarrow  y,\ s.\ t.\ \varOmega_P,
\end{equation}
or apply prior knowledge $\kappa$ to joint datasets $\chi$  to make a judgement $\jmath$,
\begin{equation}\label{kf_comp}
\jmath=\kappa(\chi), \ s.\ t.\  \varOmega_P,
\end{equation}
where $\varOmega_P$ is the secure exchange protocol that each party $P_i$ must be subject to.

It is worth emphasizing that, according to the definition, either knowledge discovery or knowledge utilization on multiple participants must be secure and privacy-preserving. That means, primitive data of each party must be kept locally and can be utilized only after encryption or embedding. Moreover, it is desired that federation should be in the performance better than only using isolated local data, and approximate to the centralized computing way.

Since it can meet the privacy protection requirement and break the dilemma of data isolation, knowledge federation has a wide range of application prospects, especially in the financial industry. For example, a consumer has the income data in a bank, and has already got two credit cards from the other two financial institutions. With his/her monthly income and granted credit data from different institutions, the accumulated credit risk can be more reliably assessed through knowledge federation. When this customer applies for a new credit card from another institution, the accumulated risk can be used to help this institution make a better decision on whether to approve this application or not. 
As strongly regulated in most countries, both income and credit limit are extremely private and need to be strictly preserved, which can be achieved in knowledge federation.

\subsection{Conceptual Architecture}


In artificial intelligence, knowledge generally involves three key elements: valuable information, models or rules, and knowledge representation. 
In fact, federation can happen at every level or multiple levels. As illustrated in Figure~\ref{fig:four-level}, according to the stage of federation, or from primitive levels and more advanced levels, we design Knowledge Federation as a four-level conceptual framework.
\begin{itemize}
	\item Low level. In this level, the federation takes place at the early stage of computing or learning, it basically assembles all isolated ciphertext after data encryption. The encryption must be homomorphic so that the following computing or learning can work normally on the encrypted space. Since each data is seperately processed into new information before federation, this level is also called information level.
	\item Middle level. When the federation occurs during model training, knowledge federation is equivalent to federated machine learning to some extent. In this level, local models are iteratively updated through aggregating models on a third-party server. Model updates are usually encrypted with such technologies as differential privacy before uploading to the server. As a consequence, model level is another name of this scenario.
	\item High level. Cognition refers to the formation of knowledge through collected data. In high level, coarse cognition is first locally extracted on each party, and then the federation works on the coarse cognition in order to produce fine cognition or meaningful knowledge. As the ensemble of local cognitions, this federation is similar to ensemble learning in some sense, but the noticeable difference between them is that the latter is irrelevant of multi-party data or privacy preserving. Cognition is so important to the high level that we also call this level as cognition level. 
	\item Top level. Once knowledge is created or learned, it will be stored in knowledge warehouse and shared with other entities. In top level, all knowledge is viewed as independent knowledge nodes that connect each other to construct knowledge network.  Roaming and exploring on the network, one can produce or infer more knowledge for decision making. This level is also called knowledge level.
\end{itemize}

\subsection{Information Level}
\label{chap:info}
\begin{figure}
	\centering
	\includegraphics[width=1.0\linewidth]{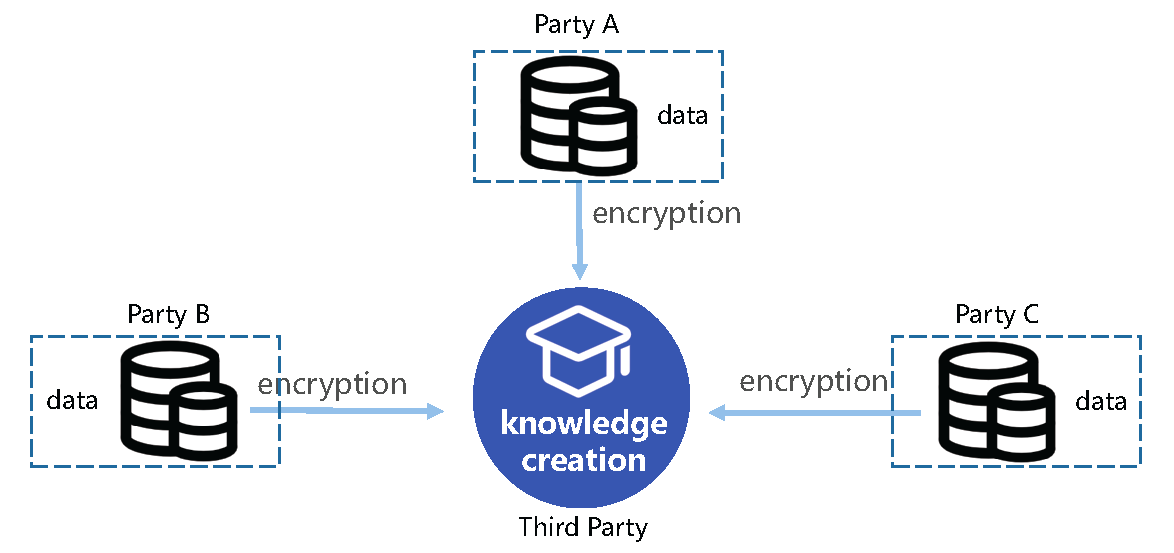}
	\caption{Information-level  knowledge federation. Original data on each party must be converted into encrypted information before uploading to the third-party server. Moreover, the knowledge creation cannot involve direct decryption of ciphertext on the server.}
	\label{fig:information}
\end{figure}

As shown in Figure \ref{fig:information},
information-level federation requires that original data must be encrypted on each party before uploading to the third-party server. It is worth noting that the encryption is supposed to be homomorphic since the following computing or learning on the server directly works on the ciphertext space whithout decryption. That is, original data 
$\chi$ is first converted into encrypted information $\chi_g$ through an encryption function
$g:\chi\rightarrow  \chi_g$, and then the significant knowledge $\kappa$ is directly produced on $\chi_g$ with labels $y$,  
$$\kappa:\chi_g\rightarrow  y_g,$$
or the reasonable judgement $\jmath$ is directly made with prior knowledge $\kappa$ on $\chi_g$,
\begin{equation}\label{kf_comp}
\jmath=\kappa(\chi_g).
\end{equation}


Information-level federation was studied previously by \cite{ML-Confi-2013} and \cite{Encrypted-2015}, where
privacy-preserving machine learning is based on
fully homomorphic encryption (FHE). Subsequently, CryptoNets, the first
neural network over encrypted data, was proposed in \cite{CryptoNets-2016} to do the inference of privacy-preserving
deep learning. 
Others cryptographic techniques were also applied in \cite{ONN-2017,Outsourced-2018} to achieve similar goals. In addition, to support both the training and inference phases, 
a CryptoNN framework was proposed in \cite{CryptoNN-2019} to train a neural network model over encrypted data by using a functional encryption scheme. \cite{LR-pack-2019} developed a privacy-preserving solution to learn regularized linear regression models using a linearly homomorphic encryption (LHE) scheme. \cite{SD-LR2019} combined differential privacy methods and homomorphic encryption techniques for logistic regression.
For more information, please refer to the survey \cite{survey-2019}.

There are some examples involving information-level federation applications, including the secure prediction of neural networks \cite{Outsourced-2018}, the secure retrieval of data from encrypted databases \cite{secure-search-2018}, 
classification \cite{classification-2019} and document ranking \cite{document-2019}. The challenge is that the current technological status and the efficiency issues still restrict the wide applicability of information-level federation in practice.

\subsection{Model Level}
\label{chap:model}

\begin{figure}[t]
	\center
	\includegraphics[width=\linewidth]{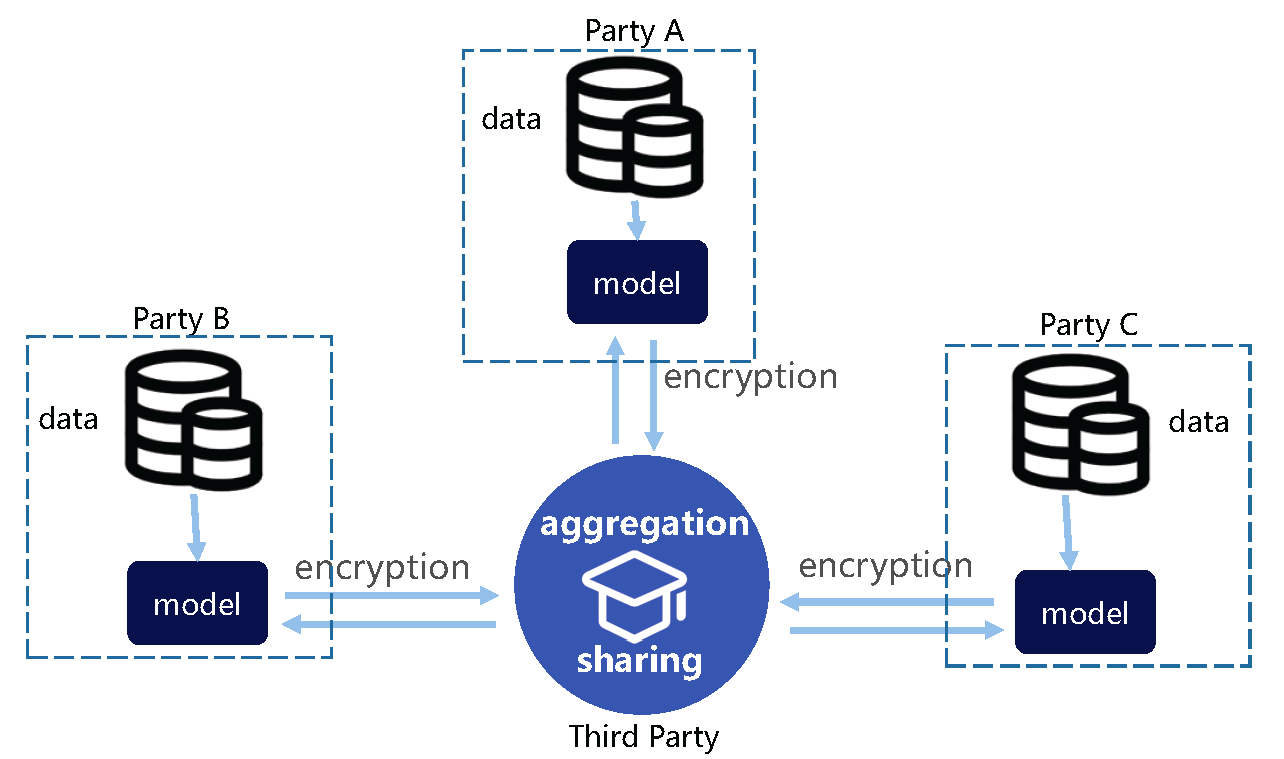}
	\caption{Model-level knowledge federation. Local models are independent trained on each party, local model updates are aggregated on the third party after encryption to generate global updates, and models are iteratively optimized through sharing global updates to all parties.}
	\label{fig:model}
\end{figure}

Model-level knowledge federation mainly concerns about how to extract global knowledge based on local models, as demonstrated in Figure \ref{fig:model}. 
The way of gathering local models varies a lot with data distribution on each party.
Taking into account the difference of data distribution in real applications, model-level federation is further classified into three types, cross-sample, cross-feature, and hybrid federation.

\subsubsection{Cross-sample Federation} 

In cross-sample cases, data with same features is distributed on each party, but samples or users on a party are independent and mostly disjoint from other parties, which is also called horizontal federated learning in \cite{yang2019federated}. Labels for samples will be collected respectively on each party. The federation aims to take  advantage of all these samples to train a common model through aggregating model updates rather than uploading local data to the centralized third-party server. Since local labels are only used to supervise local models and do not need to be transmitted among different parties, label privacy is secure as well.

Cross-sample federation is applicable for protecting user privacy on intelligent portable terminals, but it often faces the challenge of less samples for each participant, so extending few-shot learning algorithms (e.g.\cite{MAML17}) to the federated version is very valuable \cite{Fed-meta-learning}.
A typical example of cross-sample federation application is the smart keyboard app that predicts next words on mobile phones~\cite{konevcny2016federated0}.

\subsubsection{Cross-feature Federation} 	
When there exist common samples, although with different features, among several parties, fusing separate features of common samples will be helpful to model improvement. Unfortunately, it is unacceptable to directly concatenate them for the sake of data security. Cross-feature federation is a way of federated learning that both considers comprehensive features and prevents data leakage, which is called vertical federated learning in \cite{yang2019federated}. In
\cite{hardy2017private, nock2018entity}, the authors described a cross-feature federation scheme to train a logistic regression model and adopted homomorphic encryption for privacy-preserving computations.

This federation has been in demand for financial risk control since the rapid development of Internet finance. But cross feature still faces two challenges. 
One is how to prevent user privacy leakage during aligning common samples between parties. Since
there are different samples among participants, we need to find their common users before training. Gennerally, user data outside (and even within) the intersection is not expected to be known by each other.
Obviously, simple collision between databases cannot solve this problem, therefore secure data inquiry methods are needed to be elaborately designed.
The other is how to preserve label privacy while training models on the party with no labels collected.
In cross-feature federation, each participant holds part of features, but only one participant has the label as the ground truth for training. As a result, models cannot be trained independently on any participant without labels. A possible solution is to construct an oblivious transfer protocol based on all the possible labels.

\subsubsection{Hybrid Federation}
Except for cross-sample and cross-feature, there is another more complex scenario where only a small portion of samples or features are intersected among all participants. Herewith, the federation involves the hybrid of cross-sample and cross-feature, so we refer to this setting as hybrid federation. To make the best of available data, transfer learning or knowledge distillation can be used to provide federation solutions for the entire sample and feature space. The transfer federated learning method proposed in \cite{liu2018secure}  explores hidden representation of incomplete features and samples through adapting extracted knowledge to target domain.

This federation is more common but more challenging in real applications. For example, suppose there are two institutions, one
is a local insurance company located in a city, and the other is a hospital located in another city.
Obviously, only a small portion of user samples is possibly intersected between these two parties due to different geographical areas. 
Moreover, business differentiation determines that overlapping features between them is quite limited. If we expect to utilize data from these two institutions to train a model for insurance risk assessment, hybrid federation will come in handy.

However, the cross-sample, cross-feature, and hybrid federations are not only limited to the model level, but also arise on the other three levels. It is worth noting that they can be applied to other secure multi-party applications  such as secure multi-party computation, secure multi-party prediction, and secure multi-party inference.

\subsection{Cognition Level}
\label{chap:cog}
\begin{figure} [htb]
	\center
	\includegraphics[width=\linewidth]{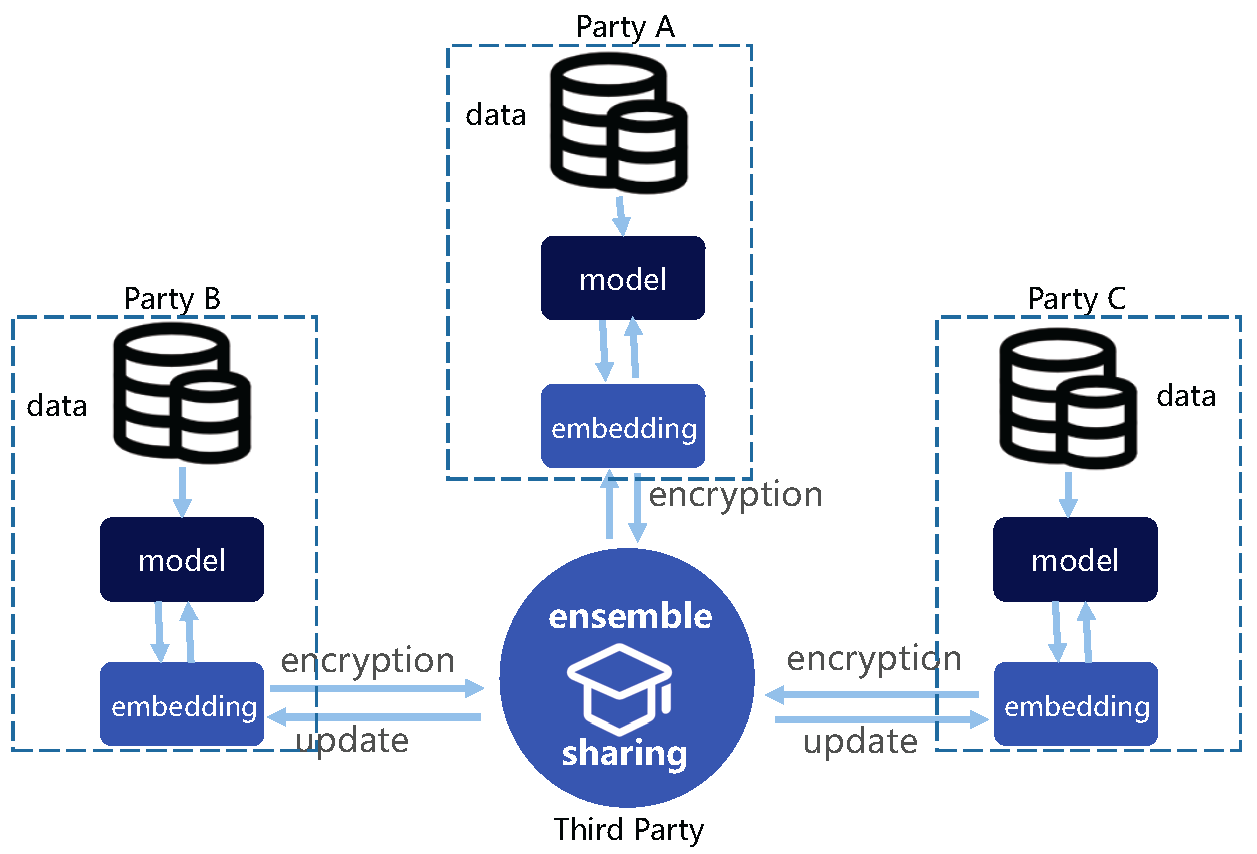}
	\caption{Cognition-level knowledge federation. Federation comes about over feature embedding and can be achieved with an ensemble model.}
	\label{fig:cognition}
\end{figure}
The obvious distinction between cognition and model levels is that embedded features, rather than model updates, will be encrypted and applied to further ensemble in cognition-level federation. The embedding could be the last fully-connected layer in deep neural networks or the local cognition extracted on a party. The ensemble on the third party is a training process with an independent model based on local embeddings, and the training will interact with local models and iterate  until convergence.

To be more specific, as shown in Figure \ref{fig:cognition}, during federation, high-level features embedded in local data are first encrypted and sent to the third-party server.  Then the server performs knowledge discovery through training an ensemble model. The ensemble model will reversely direct the optimization of local models. Local embedding can be viewed as coarse cognition (or meta knowledge) that is brought together to create fine cognition (or global knowledge). 

If each party has diverse data and a model of the same type, the cognition-level federation is similar to ensemble methods\cite{Zhou2012Ensemble} in a sense, except that the former is more concerned about privacy preserving. In fact, the selected model may vary in structure or type from one party to another. The diversity of the models that make up the federation will lead to the success of a federation system.

Pratical applications are often in need of this federation. For instance, if we want to comprehensively analyze and predict user behavior through the multi-source heterogeneous data including video, audio and text, cognition-level federation should be the best way for extracting global behavior knowledge while preserving respective data privacy.

\subsection{Knowledge Level}
Once the initial knowledge is constructed in a certain way and saved in a knowledge base or knowledge graph, the federation moves to a higher-level phase, knowledge-level federation, where initial knowledge from multiple knowledge bases will further collaborate and jointly derive new or advanced knowledge. In order to ensure that knowledge is able to flow easily among multiple parties, a knowledge network should be first constructed through connecting all knowledge nodes, each of which represents an independent knowledge base or a subgraph of knowledge.
In a nutshell, with security and privacy preserving, the knowledge-level federation expects to let knowledge freely flow in the knowledge network and deduct more comprehensive and valuable knowledge through knowledge fusion or reasoning. 

It should be emphasized that knowledge network is different from, but closely related to, knowledge graph. The latter mainly describes entities and their interrelations, organized
in a graph, as discussed in \cite{ehrlinger2016towards}. Knowledge network is built on the top of knowledge graphs, and is designed as a network of relevant knowledge in a specific context, task, or domain among to multiple organizations.  
In this case, 
knowledge fusion and reasoning techniques \cite{Grosan2011,KnowledgeFusion-2014}  can be applied to yield solutions on the network under a federation. For example, consider two pieces of knowledge, one is that 
a company has a record of tax evasion in a knowledge node in Party A; and the other is that this company is unable to offset debts with assets in another node in Party B. With other knowledge of contexts and banking policies, credit risk model can thus be comprehensively assessed through the knowledge-level federation.

\section{Unification of Secure Multi-party Computation and Learning}

Knowledge federation is a unfied framework for secure multi-party computation (MPC) and multi-party learning (MPL) since the computing or learning task can be achieved respectively under this framework. The notable difference between MPC and MPL is that the latter requires to train a model with the multi-party input data, but the former does not. In the knowledge federation framework, both MPC and MPL are unified as the federation process that takes place on a virtual or physical, but always independent, third-party server.

\subsection{Secure Multi-party Computation}
In knowledge federation, MPC is implemented with the help of a weakly centralized third-party server that is only responsible for computing and will not store data persistently.
In this case, the existing knowledge is squarely utilized by performing homomorphic operations  on the server, which requires that data must be homomorphically encrypted \cite{rivest2018on} before uploading. 
Differential privacy (DP) can also be used in secure MPC by adding extra noise for the plaintext \cite{geyer2017differentially}.  However, DP methods can only deal with simple operations, thus limit their application scope.

Since the operation runs on the ciphertext space, this computing procedure will not leak data privacy, even if the third party is probably untrustable. 
If the third-party is virtual or omitted, the proposed framework is fully decentralized, which is quite useful especially in two-party collaboration.
According to the forementioned description in Section \ref{chap:info}, secure MPC is radically a special case of the information-level federation with regard to computation.

\subsection{Secure Multi-party Learning}
In some situations, the knowledge must be jointly learned on input data from each party, where it is in essence a secure multi-party learning (MPL) problem. 
There are three ways of secure federation in MPL. 

A natural idea is that local data is first homomorphically encrypted and then sent to the third-party server, models are trained on the server with conventional machine learning or deep learning methods. This case is classified as the information-level federation in the KF framework.

Another popular way of implementing secure MPL is often called federated learning (FL) in literature~\cite{konevcny2016federated0,yang2019federated,kairouz2019advances}. A model is first locally trained with isolated data on each party, model updates are then gathered and computed on the third party, and the model is finally updated and trained iteratively in this way until convergence. This training method is classified as the model-level KF introduced in Section \ref{chap:model}. 

MPL can also be achieved through the ensemble of embedded features on the third-party server. The ensemble procedure essentially amounts to training a comprehensive model over multi-party data. Data security and privacy can be preserved with homomorphic encryption on local embeddings. This way is equivalent to the cognition-level knowledge federation in Section \ref{chap:cog}.

Beyond the focus of secure multi-party computation and learning, the KF framework includes also secure multi-party inference and sharing that may involve securely fusing encrypted data and knowledge from multiple parties in nontrivial manners. 
The advantage of the unified KF framework is that one platform can integrate various federated applications during the reference implementation, avoiding unnecessary trouble and chaos that developing and applying multiple platforms will cause.

\begin{figure}[ht]
	\center
	\includegraphics[width=1\columnwidth]{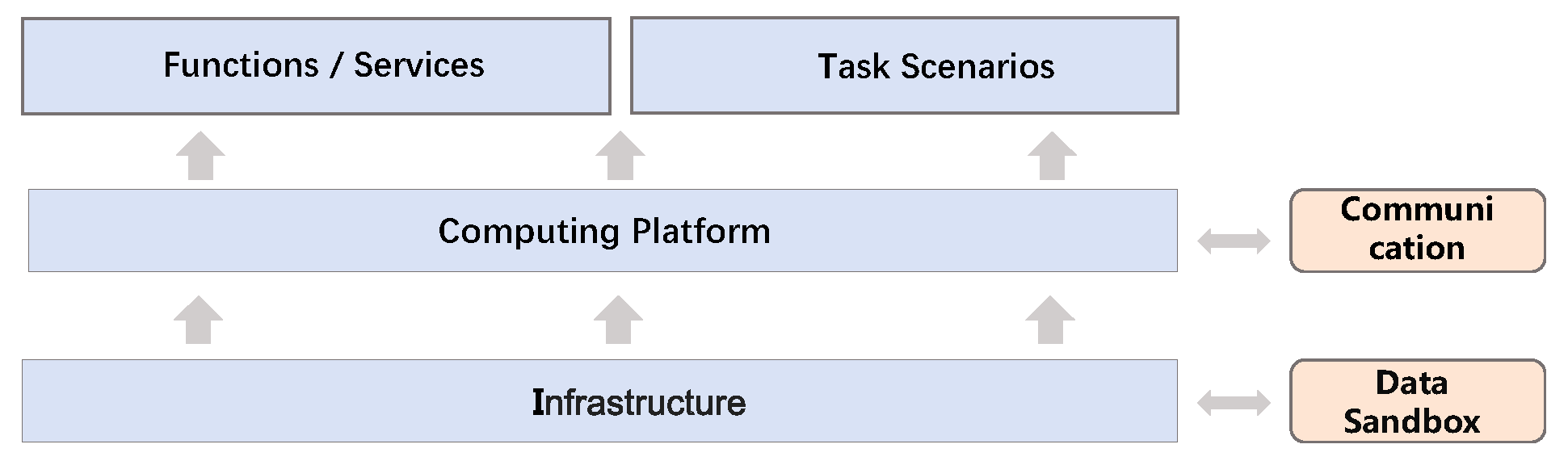}
	\caption{Overview of iBond Platform.}
	\label{fig:iBond_arch}
\end{figure}

\section{iBond Platform: a Reference Implementation of Knowledge Federation}

\subsection{Platform Overview}
Following the guideline of Knowledge Federation (KF), iBond platform unifies secure multi-party computation, learning, inference, and reasoning, all with data security and privacy preserving. As shown in Figure~\ref{fig:iBond_arch}, iBond platform consists of four core modules and two middlewares. 

Core modules are the important and necessary components that implement key functions to the platform, specifically including:
\begin{itemize}
\item Functions/Services. This module provides common functions and services such as account management, data management, model management, expense calculation, contribution assessment  etc, and well-defined external interfaces capable of utilizing these services.
\item Task Scenarios. For such task scenarios as credit score and multi-party loan,  several models or strategies have been designed and integrated into the platform by default. When a new task arises from a particular application scenario, a new model or knowledge is needed to be customized and developed by algorithm engineers.
\item Computing Platform. This is an open pool of computing resources in order to complete the federalization of traditional algorithms from many areas, such as NLP, image classification, and machine learning. In addition, it supports data encryption and decryption methods as well.
\item Infrastructure. Here are all basic facilities needed in the iBond platform. This module covers utilities from task monitoring, resource scheduling, data access, to model storage.
\end{itemize}

In the iBond platform, middleware is an abstraction layer that hides detail about hardware devices or other software from core modules, makeing it easier to implement communication and data standardization import. Two middlewares are repspectively described as follows:
\begin{itemize}
\item \emph{Communication}. This layer is mainly in charge of connecting Intranet or Internet with the platform. Meanwhile, it can join the platform to test and production environments for rapid model deployment. 
\item \emph{Data sandbox}. In the sandbox, data is first loaded from multi-source heterogeneous databases, then normalized and standardized according to a specified rule, and finally transformed through de-identification, desensitization, and encryption for the purpose of data security and privacy preserving. 
This layer also encapsulates some secure data exchange protocols for multi-party federation. The major concern is secure query/retrieval, secure sample alignment, secure aggregation, etc.
\end{itemize}

\begin{figure*} [ht]
	\center
	\includegraphics[width=1.8\columnwidth]{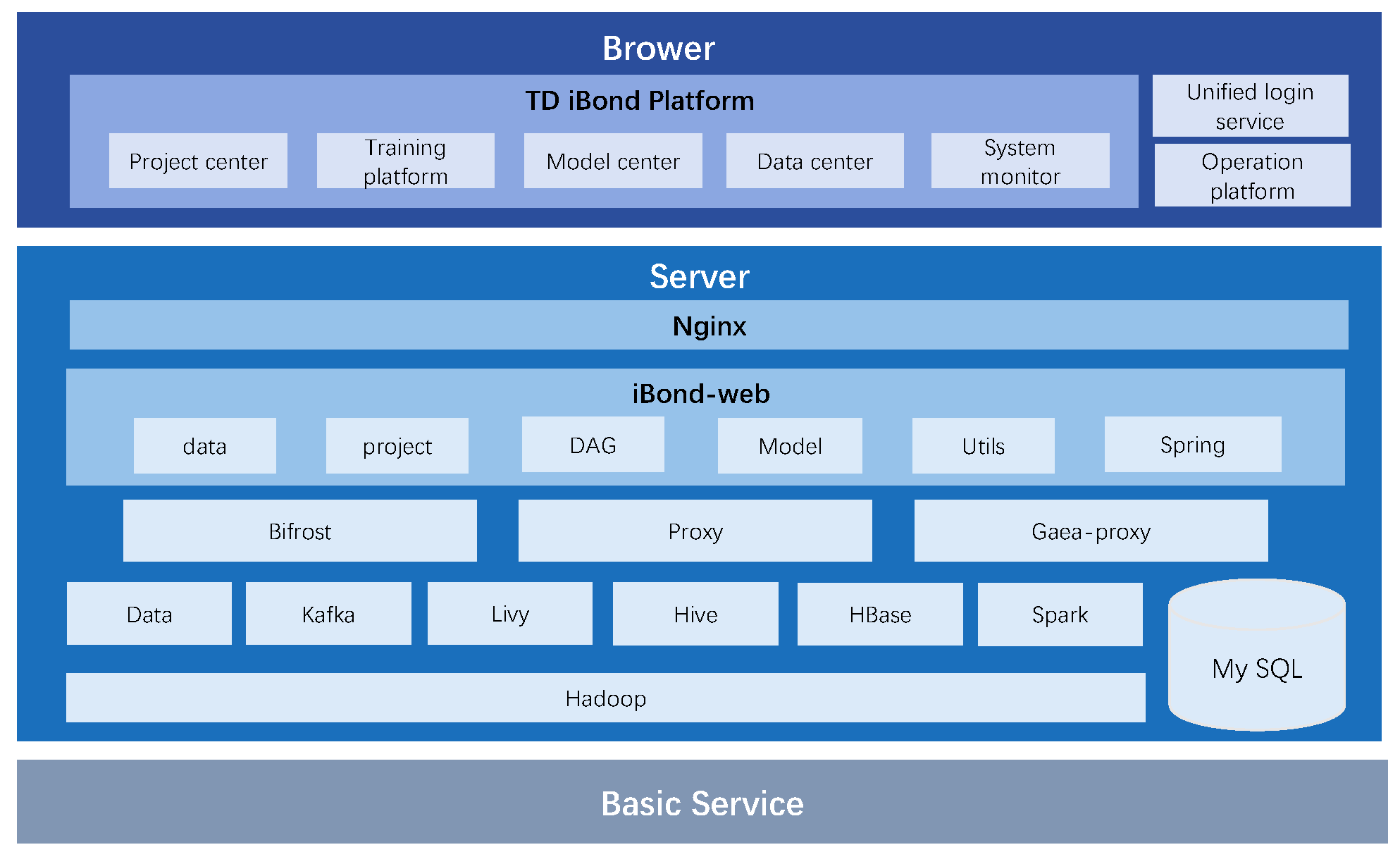}
	\caption{The architecture of iBond platform.}
	\label{fig:iBond}
\end{figure*}

\subsection{Platform Architecture}
\begin{figure*} [ht]
	\center
	\includegraphics[width=1.8\columnwidth]{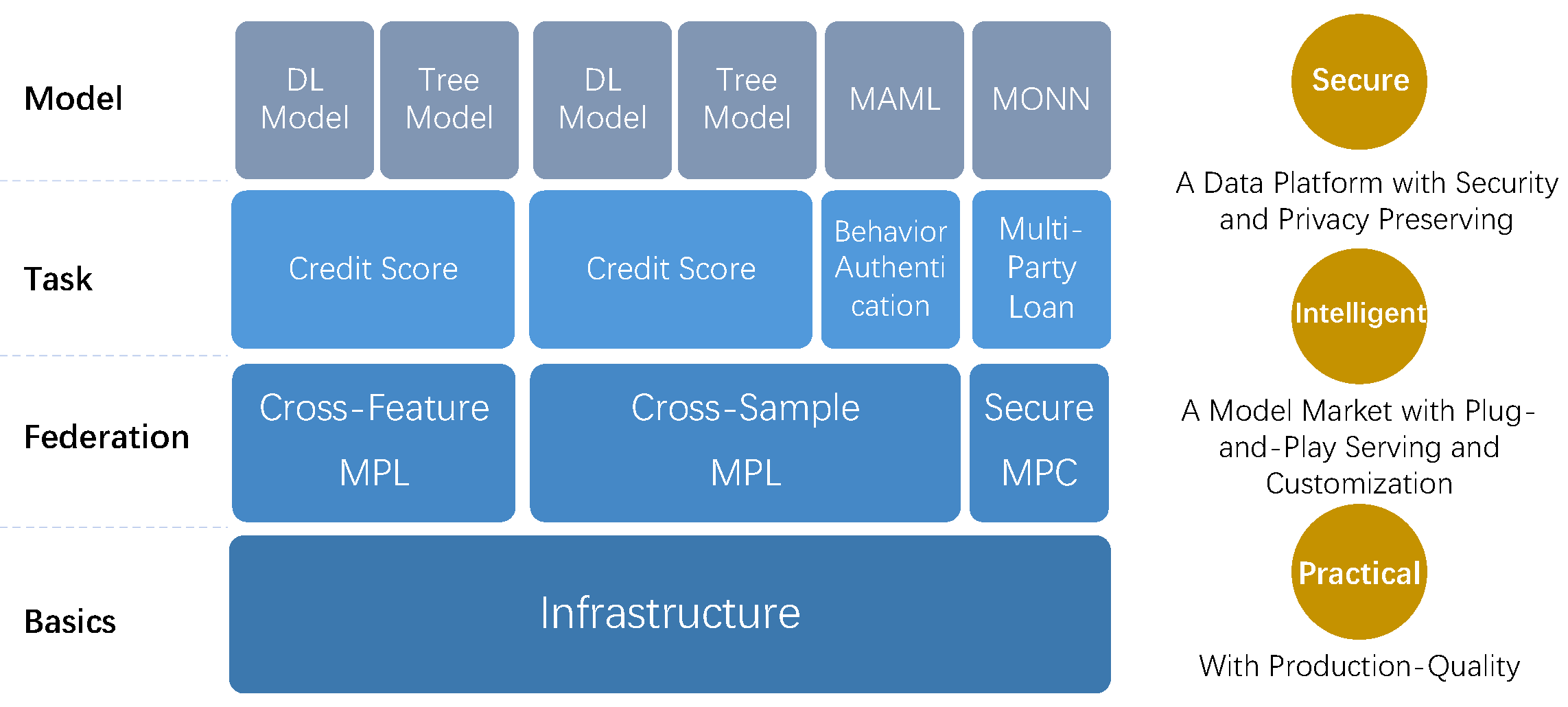}
	\caption{The illustration of the setting customized for financial applications.}
	\label{fig:fin}
\end{figure*}

Figure~\ref{fig:iBond} illustrates the architecture of iBond Platform. 
One of the advantages of this platform architecture is that it is very convenient to transplant the platform from one computing environment to another for deployment and implementation, where the containerization technology is used to package up code and all its dependencies. 
Figure~\ref{fig:fin} illustrates the special customization for financial applications. Multiple intelligent models have been desinged and preseted in iBond platform for different tasks such as credit score, behavior authentication and multi-party loan, and changing federation manners from cross-feature, cross-sample MPL to secure MPC.

It is worth highlighting that three features of iBond are respectively, 
\begin{enumerate}
	\item Secure. It is a data platform for federation with security and privacy preserving.
	\item Intelligent. It supports a model market where intelligent models are plug-and-play and customizable.
	\item Practical. It is practical and production-quality with sufficient functions in order to support the complete MPL or MPC processes from sample alignment until model deployment.
\end{enumerate}

In addition to key components in computing, learning, and reasoning, iBond supports the third-party (could be virtual or physical) to act as an arbitrator for assessment, regulation, and arbitration to meet the requirements in many data-sensitive applications, such as finance, health, insurance, and government. The third-party role is facilitated with many utilities for visualization, data flow tracking, deep packet inspection, and assessing potential data leakage and privacy break. More details can be found in our recently published white paper \cite{KF-whitepaper}.

\subsection{Security and Privacy Concerns}

The iBond platform takes into comprehensive consideration of both data security from the whole process of federation, and privacy preserving with de-identification and desensitization technolgies. In particular, the followling security measures are worth emphasizing, 
\begin{enumerate}
	\item Data Normalization and Standardization. Local data in each party is loaded to and preprocessed by a sandbox. All data are then standardized with consistent de-identification and desensitization methods. The output from the sandbox is in the uniform naming format and measuring unit. 
	\item Secure Data Exchange. Secure data exchange protocols, including sample alignment, data query, model aggregation, are implemented in the data sandbox so as to protect data security and privacy. Common cryptography algorithms such as homomorphic encryption, secure key exchange, one-time-padding, and secret sharing are offered to customize other secure data exchange protocols.
	\item Trustable Third-Party. The third party in iBond architecture acts as an arbitrator and focuses only computation, it will never store local data. As a result, the computing procedure over the third party will not leak data privacy, even if the third party is untrustable. 
	\item Network Transmission Security. The overall network is divided into three security domains according to the security level: i) private domain for local data, ii) exchange domain for exchangeable models or strategies, iii) federated domain for arbitrator. In addition, the connection between participants and arbiter adopts the secure transmission channel through asymmetric encryption and bidirectional authorization.
\end{enumerate}

\subsection{Application}

\begin{figure*} [th]
	\center
	\includegraphics[width=1\columnwidth]{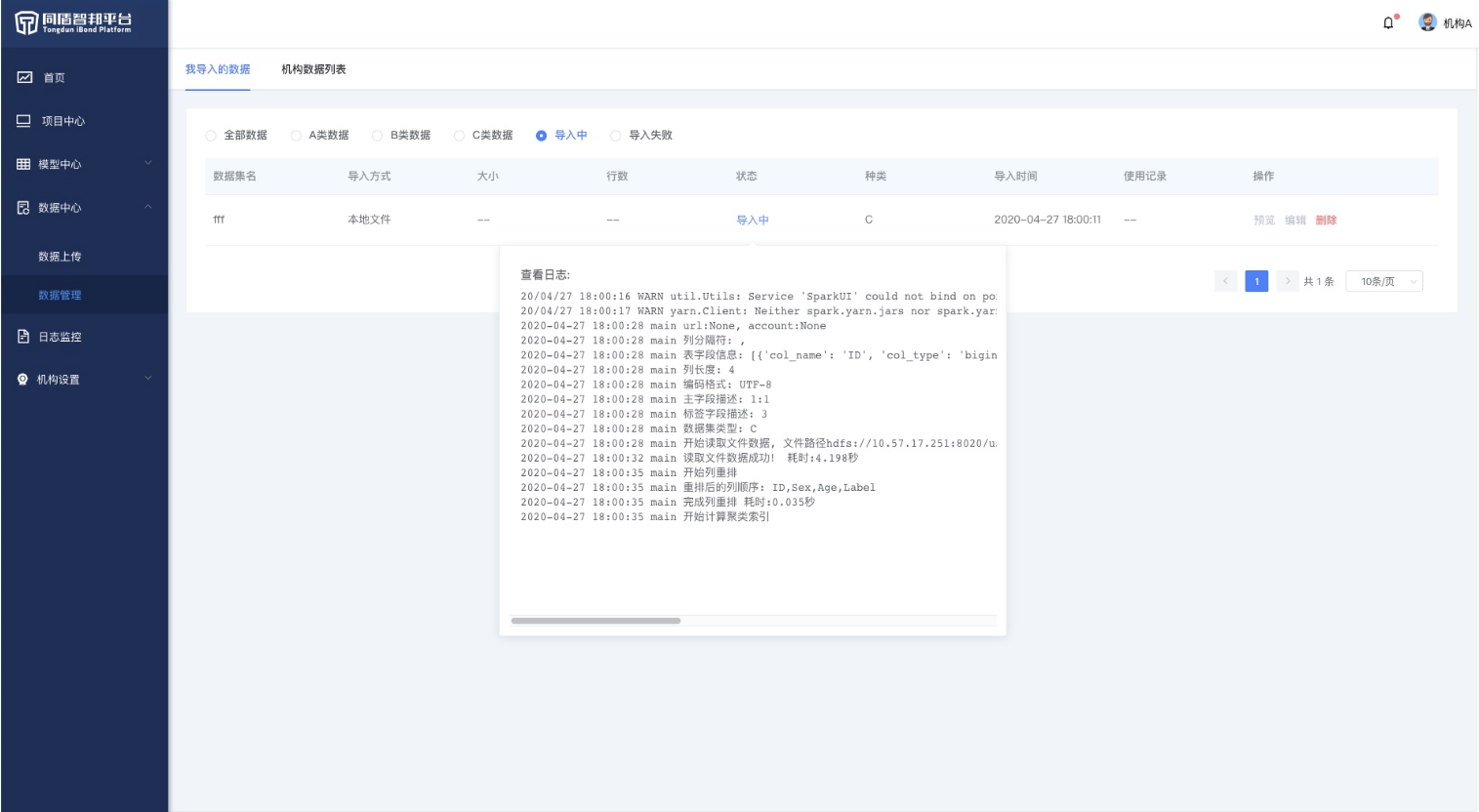}
	\includegraphics[width=1\columnwidth]{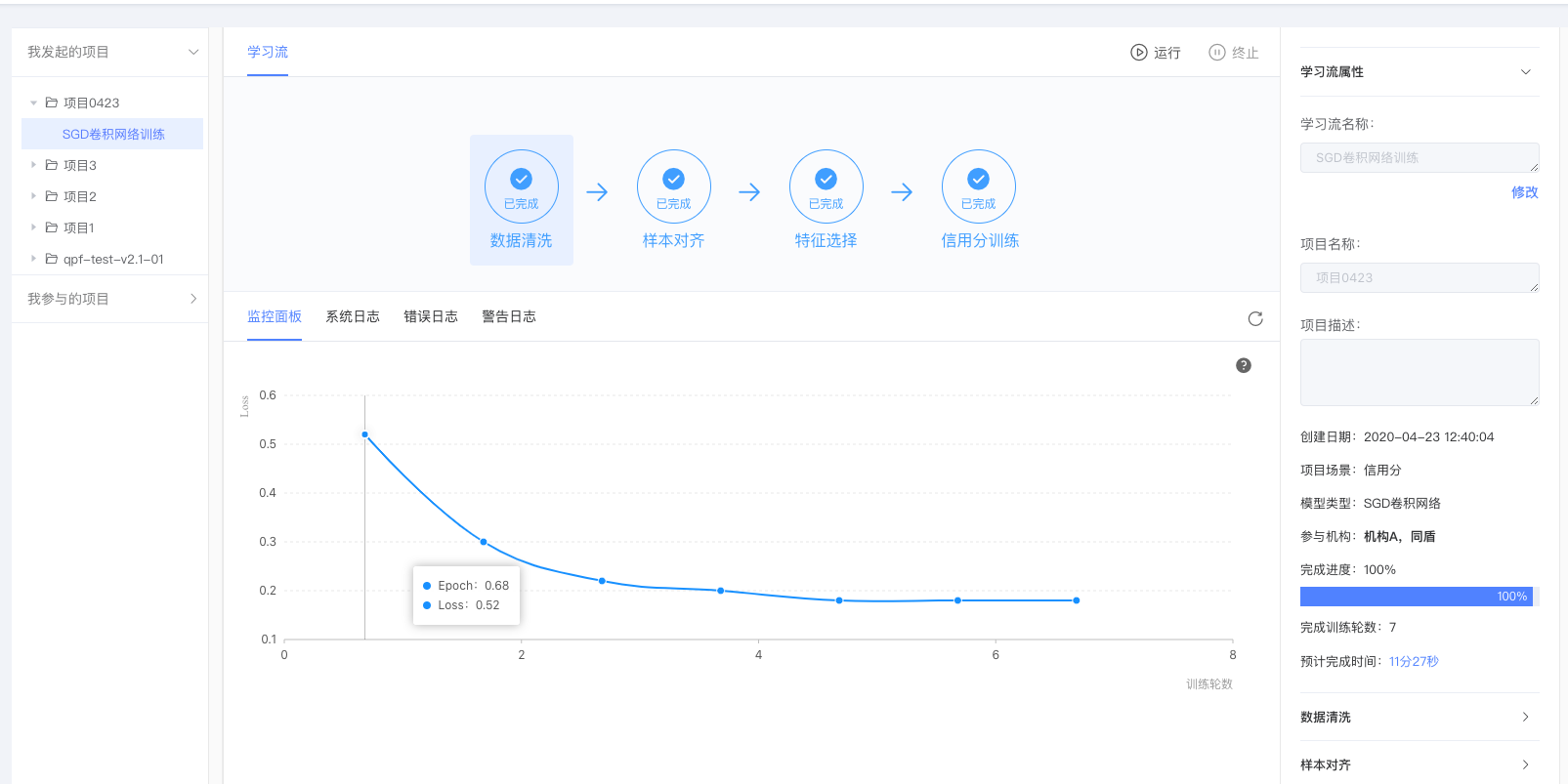}
	\caption{Interactive interface of iBond platform. Left: data import, Right: loss curve.}
	\label{fig:app}
\end{figure*}

In this section, we present a practical application scenario involving credit score assessment in financial indusry. Suppose that there are two parties with heterogeneous features, only one party has labels, and they want to jointly train a credit score model. This is actually a cross-feature MPL problem. The third party is assumed to be a virtual server to act as an arbitrator. Based on iBond platform, 
the connection between both parties and the arbitrator can be established for federation. 

Aftern local data on both parties is loaded and standardized as shown in the left of Figure~\ref{fig:app}, secure sample alignment can be launched to find the intersection set in terms of a specified identifier, where those different samples outside the set are unknown to each other. 
Subsequently, secure feature selection can start to work for picking up the features significant to credit score model. 
Finally, the federated model for credit score can be created and trained until convergence. In the case, we choose a deep neural network, specially designed for cross-feature federation in the platform, as the credit score model. 
During training, model updates are safely gathered on the third party based on the secure model aggregation protocol. 
The loss curve is presented in the right of Figure~\ref{fig:app}.
In order to ensure that labels are kept locally and will not leak out, the platform provides a protocol of secure label transfer, in which the party with no labels will calculate gradients for all possible labels and transfer them to the other party with labels for selection and aggregation. 

In sum, iBond offers a production-quality KF platform and can be applied to many industrial areas with sensitive data other than finance, for example health, insurance, marketing, and government.

\section{Conclusion}

To address data security and privacy issues in conventional machine learning and deep learning applications that involve multiple parties, we proposed a unified and hierarchical framework Knowledge Federation (KF) and its reference implementation iBond Platform. KF unifies secure multi-party computation and learning in a coherent and hierarchical manner with four levels of abstractions: information level, model level, cognition level, and knowledge level. With the unified framework, KF fills the gap to meet comprehensive needs from low-level to high-level knowledge sharing, computing, and learning (discovery, representation, and reasoning) in production environments in broad application scenarios, such as finance, health, insurance, marketing, and government with sensitive data. Leveraging both secure multi-party computation, learning, sharing, inference, and reasoning, KF connects the dots scattered in multiple-disciplinary areas such as cryptography, machine learning, deep learning, federated learning, big data, and other AI techniques. The iBond Platform has been deployed in production environments serving many business customers. 

Our ongoing work focuses on an in-depth assessment of contribution models, attack models, and regulation models that facilitate regulation and arbitration during the secure data transactions and learning processes. 
It is expected that in the near future, knowledge federation would break the barriers between institutions and establish a model market where knowledge could be created and shared jointly at liberty, while preserving data security and privacy. This work also calls for the community to join together to establish a consortium of knowledge federation to push forward the best practices and design for privacy-preserving AI ecosystems.

\section*{Acknowledgement}
The comprehensive iBond platform has been developed by dozens of colleagues. We deeply appreciate their unique contributions and great efforts in numerous ways in algorithms, systems, testing, and commercial deployment and production serving.

\bibliographystyle{IEEEtran}

\end{document}